\documentclass[preprint2]{aastex_mod}
\usepackage{ulem}
\usepackage[dvips]{color}
\usepackage{graphicx}
\usepackage{txfonts}
\usepackage{rotating}
\usepackage[USenglish]{babel}

\newcommand{\oc}{$\omega$~Cen}
\usepackage{longtable}

\begin{document}

\title{A DOUBLE WHITE-DWARF COOLING SEQUENCE IN $\omega$
  CENTAURI$^{\ast}$}

\author{A.\ Bellini\altaffilmark{1},
J.\ Anderson\altaffilmark{1},
M.\ Salaris\altaffilmark{2},
S.\ Cassisi\altaffilmark{3},
L.\ R.\ Bedin\altaffilmark{4},
G.\ Piotto\altaffilmark{5,4},
and P. Bergeron\altaffilmark{6}
}

\altaffiltext{1}{Space Telescope Science Institute, 3700 San Martin
  Dr., Baltimore, MD 21218, USA}

\altaffiltext{2}{Astrophysics Research Institute, Liverpool John
  Moores University, Twelve Quays House, Egerton Wharf, Birkenhead,
  CH41 1LD, United Kingdom}

\altaffiltext{3}{Istituto Nazionale di Astrofisica, Osservatorio
  Astronomico di Collurania, via Mentore Maggini, I-64100 Teramo,
  Italy}

\altaffiltext{4}{Istituto Nazionale di Astrofisica, Osservatorio
  Astronomico di Padova, v.co dell'Osservatorio 5, I-35122, Padova,
  Italy}

\altaffiltext{5}{Dipartimento di Fisica e Astronomia ``Galileo
  Galilei'', Universit\`a di Padova, v.co dell'Osservatorio 3,
  I-35122, Padova, Italy}

\altaffiltext{6}{D\'epartement de Physique, Universit\'e de
  Montr\'eal, C.P. 6128, Succ. Centre-Ville, Montr\'eal, Qu\'ebec H3C
  3J7, Canada}

\altaffiltext{$^\ast$}{Based on archival observations with the
  NASA/ESA \textit{Hubble Space Telescope}, obtained at the Space
  Telescope Science Institute, which is operated by AURA, Inc., under
  NASA contract NAS 5-26555.}

\email{bellini@stsci.edu}

\date{Submitted on April 4, 2013, Accepted on May 1, 2013}

\begin{abstract} {
We have applied our empirical-PSF-based photometric techniques on a
large number of calibration-related WFC3/UVIS UV-B exposures of the
core of \oc, and found a well-defined split in the bright part of the
white-dwarf cooling sequence (WDCS).  The redder sequence is more
populated by a factor of $\sim 2$.  We can explain the separation of
the two sequences and their number ratio in terms of the He-normal and
He-rich subpopulations that had been previously identified along the
cluster main sequence. The blue WDCS is populated by the evolved stars
of the He-normal component ($\sim 0.55 {\mathcal{M}_{\odot}}$ CO-core
DA objects) while the red WDCS hosts the end-products of the He-rich
population ($\sim 0.46 {\mathcal{M}_{\odot}}$ objects, $\sim 10$~\%
CO-core and $\sim 90$~\% He-core WDs).  The He-core WDs correspond to
He-rich stars that missed the central He-ignition, and we estimate
their fraction by analyzing the population ratios along the cluster
horizontal branch.  }
\end{abstract}

\keywords{
globular clusters: individual (NGC 5139) ---
Hertzsprung-Russell and C-M diagrams --- 
stars: Population II --- 
white dwarfs --- 
techniques: photometric}

\maketitle

\begin{figure*}[!t]
\centering
\includegraphics[width=16cm]{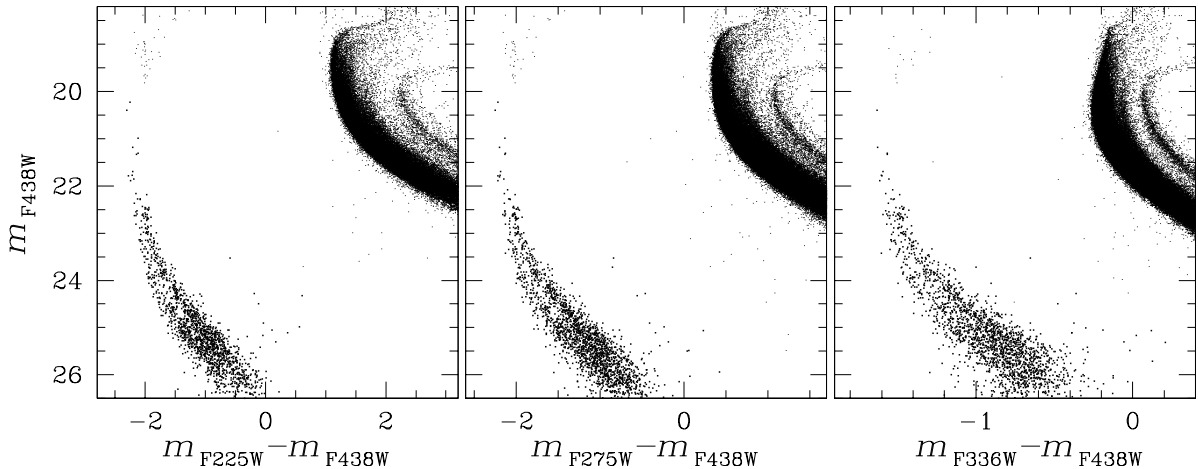}
\caption{\small{The double WDCS of \oc: from left to right we show
    $m_{\rm F438W}$ versus $m_{\rm F225W}-m_{\rm F438W}$, $m_{\rm
      F275W}-m_{\rm F438W}$, and $m_{\rm F336W}-m_{\rm F438W}$,
    respectively.}}
\label{f1}
\end{figure*}

\begin{figure*}[!th]
\centering
\includegraphics[width=16cm]{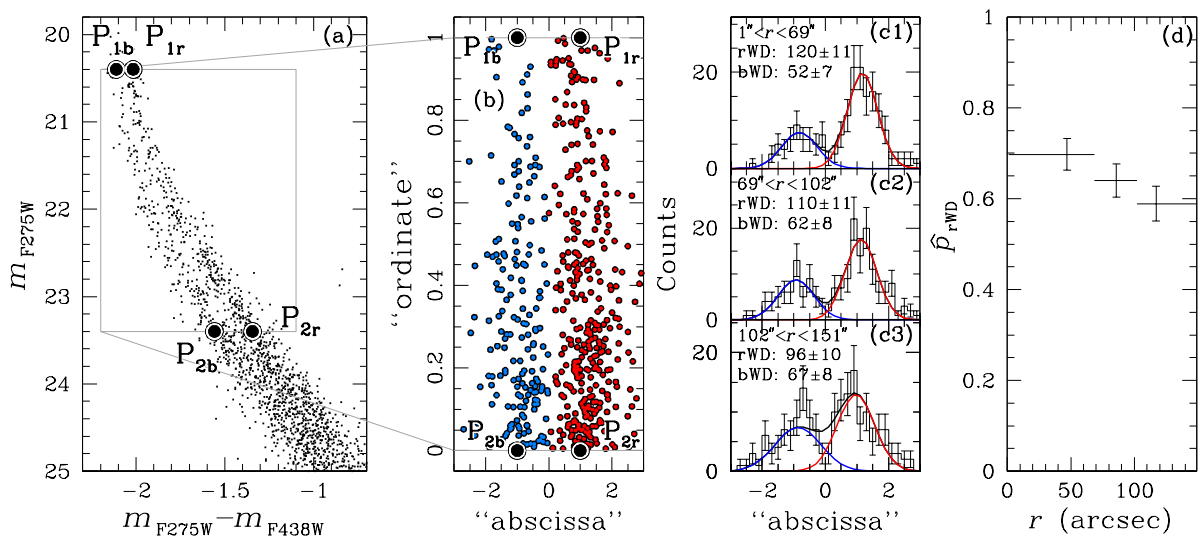}
\caption{\small{
         (a) Blow-up of the WDCS in the $m_{\rm F275W}$ versus $m_{\rm
      F275W}-m_{\rm F438W}$ CMD with the four points used in the
    coordinate-system transformation of the next panel.
         (b) Rectification of the WDCS region that was shown in gray
    in panel (a). Here we define as bWD (in blue) those stars having
    ``abscissa''$<$0, rWD stars the others (in red).
         (c) The FoV is divided into three equally-populated radial
    intervals. For each of them we fit a dual Gaussian to the
    ``abscissa'' distribution, and used the area of the two Gaussians
    to estimate the number of bWD and rWD stars in each radial
    interval (radial extension and areas are quoted in each panel).
         (d) Radial distribution of the ratio of rWD with respect to
    the total WD population.  }}
\label{f3}
\end{figure*}

\section{Introduction}
\label{sec:1}

The massive globular cluster (GC) \oc\ has been studied since the `60s
because of its peculiar stellar populations (see van Leeuwen, Hughes
\& Piotto 2002 for a review). The discovery of its multiple stellar
generations (Bedin et al.\ 2004, Bellini et al.\ 2010) has further
revitalized detailed photometric and spectroscopic investigations of
this object.  The white-dwarf (WD) cooling sequence (CS) is the
least-studied part of the color-magnitude diagram (CMD) of this
cluster, for the obvious reason that WDs are faint and are usually
concentrated in the most-crowded central regions.

Ortolani \& Rosino (1987) were the first to detect a dozen WDs in \oc,
but it is only with the advent of \textit{HST} that it became possible
to analyze the WDCS in more detail (Monelli et al.\ 2005; Calamida et
al.\ 2008).  In particular, Calamida et al.\ showed that the WDCS
spread in color is larger than expected from photometric errors alone.
We will show here that the WDCS of \oc\ actually splits into
\textit{two distinct sequences}, which can be explained as separate WD
populations of different mass and chemical stratification,
corresponding to the evolved components of the separate subpopulations
with differing helium abundances, as revealed by the photometric
analysis of main-sequence stars (e.g.  King et al.~2012 and references
therein).

\section{Data sets and reduction}
\label{sec:2}

This analysis is based on \textit{HST}'s WFC3/UVIS observations taken
between 2009 and 2011 under WFC3 calibration programs.
Specifically:\ we have $27 \times 900\,$s exposures in F225W, $31
\times 800\,$s in F275W, $37 \times 350\,$s in F336W, and $34 \times
350\,$s in F438W.  The field of view (FoV) is centered on the
cluster's center, and covers the inner $\sim 150^{\prime\prime}$.

All WFC3/UVIS \texttt{\_flt} images for a given filter were analyzed
simultaneously to generate an astrometric and photometric catalog of
stars in the field, using an evolution of the software tools described
in Anderson et al.\ (2008).  Briefly, we used information from all the
available exposures for a given filter to find significant detections
across the FoV.  These detections were then fit independently in each
exposure by employing a fixed array of $7\times8$ spatially-varying
static library PSF models, plus a $5\times5$ array of perturbation
PSFs tailored to fit the star profiles in each exposure in order to
account for focus/breathing variations.  Positions were corrected for
geometric distortion according to the routines in Bellini, Anderson \&
Bedin (2011).  The photometry was calibrated into the Vega-mag flight
system following Bedin et al.\ (2005), using the WFC3 zero points
provided by the
STScI\footnote{\url{http://www.stsci.edu/hst/wfc3/phot\_zp\_lbn}.}.

\section{The Double WD CS}
\label{sec:3}

Figure~1 shows the WDCS of \oc\ in three different CMDs, keeping the
$m_{\rm F438W}$ magnitude on the vertical axis, and using a variety of
color combinations.  WDs are shown with heavier dots, for clarity.  We
identified over 2,000 WDs across the FoV.  The WDCS can be followed
for over 6 magnitudes in $m_{\rm F438W}$, starting from just below the
termination point of the extreme horizontal branch (HB).  Two distinct
and well defined WD sequences are visible in the bright part of the
CS, before becoming indistinguishable below $m_{\rm F438W}\sim 25$.  A
few stars, about 1 magnitude redder than the WDCS, define an
additional, scarcely populated sequence.

Stars on the main sequence (MS) of \oc\ are well known to be separated
into three main components: (1) a He-rich bMS, (2) a He-normal,
[Fe/H]-poor rMS, and (3) a [Fe/H]-rich MS-a (Anderson 1997; Bedin et
al.\ 2004; Piotto et al.\ 2005; Bellini et al.\ 2010). The bMS has
been found to be more centrally concentrated than the rMS (Sollima et
al.\ 2007; Bellini et al.\ 2009), a fossil signature of the formation
process of the two populations. Indeed, the present-day relaxation
time ($t_{\rm relax}=12.3\,$Gyr, Harris 1996, 2010 edition) suggests
the two populations have not yet had enough time to mix.

In order to look for a connection between the split WDCS and the split
MS, we measured the radial distribution of the two WD components as
follows.  We started by selecting stars in the $m_{\rm F275W}-m_{\rm
  F438W}$ CMD (Figure 2a).  To optimally separate the populations, we
used a procedure developed by Milone et al.\ (2009).  We first
selected by hand two points (${\rm P}_{\rm 1b}$,${\rm P}_{\rm 2b}$) on
the blue CS, and two points (${\rm P}_{\rm 1r}$,${\rm P}_{\rm 2r}$) on
the red one.  We then linearly transformed the CMD into a reference
frame in which the coordinates of ${\rm P}_{\rm 1b}$, ${\rm P}_{\rm
  2b}$, ${\rm P}_{\rm 1r}$, and ${\rm P}_{\rm 2r}$, map to ($-$1,1),
($-$1,0), (1,1) and (1,0), respectively (Figure~2b). Here the blue
component (the bWD) maps to negative ``abscissa'' values, and the red
component (the rWD) to positive ``abscissa'' values (color coded in
blue and red, respectively, in panel~(b)).  To examine the radial
distribution we followed the methods by Bellini et al.\ (2013).  We
divided the FoV into three equally-populated radial intervals, and fit
in each of them a Gaussian to each of the color-like distribution of
bWD and rWD stars (panels~(c)).  (A dual-Gaussian fit is more robust
than simply adding up stars in histogram bins.)  The radial coverage
of interest and the areas of each of the two Gaussians are given in
the top-left corner of each of these panels.  Panel~(d) shows the
quantity $\hat{p}_{\rm rWD}=N_{\rm rWD}/(N_{\rm rWD}+N_{\rm bWD})$ as
a function of distance from the cluster center. Binomial errors are
$\sim 3.7$~\%.  There is marginal evidence for a gradient, with rWD
stars more concentrated that bWD ones.  Moreover, rWD stars (326) are
$\sim 64$~\% of the total bWD+rWD stars (507).  We note that Bellini et
al.\ (2010) also did not find any strong evidence of a radial gradient
in the ratio of the two main MSs in the inner 2 arcmin of \oc\ (see
their Fig.~18).

The top and middle panels of Fig.~3 show the WDCS CMD in a wide
variety of color systems, with the same stars color-coded the same
ways in all the panels.  The split is present in all of them.  It is
interesting to note, however, that even though all the other
populations (MS, sub-giant branch, etc) show dramatic splits in
two-color diagrams (2CDs), there is {\it no\,} split in the 2CDs
plotted in the bottom panels of Fig.~3.  This implies that the split
is likely due to a discrete mass/radius difference between the
populations, rather than an atmospheric effect.

\begin{figure}[!ht]
\centering
\includegraphics[width=\columnwidth]{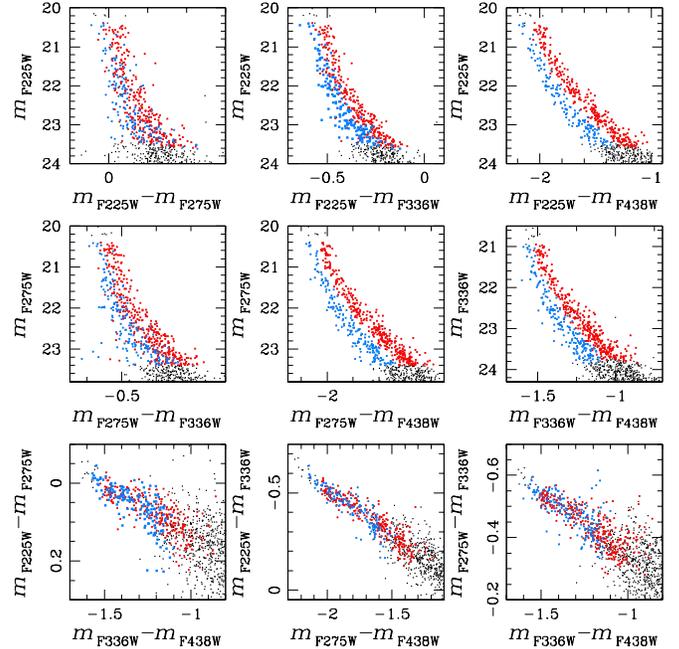}\\
\caption{\small{\textit{Top and middle panels}: multicolor analysis
                        for the WDCS. 
                \textit{Bottom panels}: two-color diagrams. Red and
                        blue WDs are those defined in Figure~2b.}}
\label{f2}
\end{figure}

\section{The nature of the two WDCSs}
\label{sec:4}

We have adopted the distance modulus derived from the cluster
eclipsing binary OGLEGC~17 (Thompson et al.~2001), i.e., $ (m-M)_{\rm
  o}=13.66\pm0.12$, and a mean value of the reddening $E(B-V)=0.12$ as
in King et al.~(2012).  To calculate the extinction in our filters we
have applied the extinction law of Cardelli et al.~(1989, with $R_V$ =
3.1) to our WD spectra used for the calculation of the bolometric
corrections to the UVIS filters
\footnote{\url{http://www.astro.umontreal.ca/\~{}bergeron/CoolingModels}.}.
The resulting extinctions in the F275W, F438W and F336W filters are:
0.75, 0.50 and 0.62~mag, respectively.  These values were applied to
our WD tracks -- transformed to the UVIS filters -- from BaSTI CO-core
WD models by Salaris et al.~(2010), He-core tracks by Bedin et
al.~(2008a) and Serenelli et al.~(2002), and additional CO-core models
calculated specifically for this project.  We focus here on the
$m_{\rm F275W}$ versus $m_{\rm F275W}-m_{\rm F438W}$ CMD, which shows
a very clear separation of the two WDCSs.

The brightest few magnitudes of a WDCS are populated by objects that
just entered the cooling sequence and are therefore cooling extremely
rapidly.  We can thus approximate all the stars in this transient
stage as having the same WD mass, since they presumably all started
with the same turnoff (TO) mass.  It is therefore also safe to match
them with single-mass cooling tracks rather than formal WD isochrones.
Theory predicts a typical WD mass of $\sim 0.54 {\mathcal{M}_{\odot}}$
in old populations (see, e.g., Weiss \& Ferguson~2009), which is
consistent with what Kalirai et al.~(2004) found in a spectroscopic
study of six bright DA WDs in M~4.

\begin{figure}[!ht]
\centering
\includegraphics[width=\columnwidth]{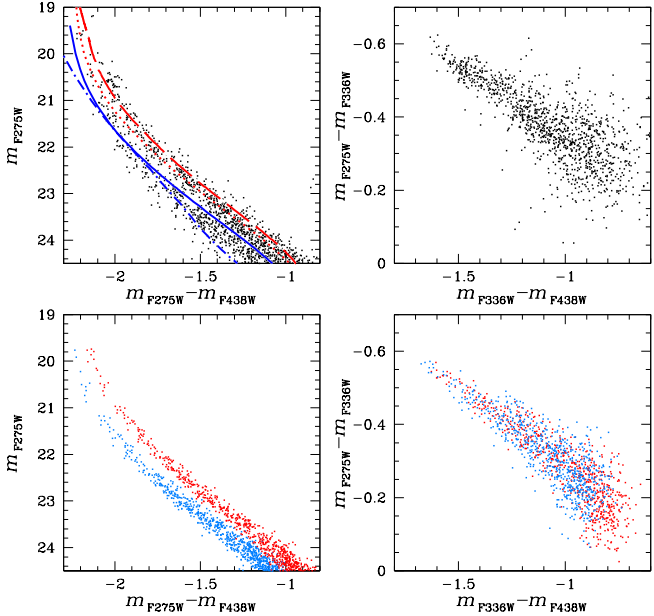}
\caption{\small{
   \textit{Left panels}: one of the observed CMDs of Fig.~\ref{f2}
   (top) and a theoretical counterpart (bottom). WD tracks for
   CO-core 0.55${\mathcal{M}_{\odot}}$ DA (blue solid) and DB models
   (blue dash-dotted), 0.46${\mathcal{M}_{\odot}}$ DA CO-core (red
   dotted) and He-core (red dashed) models are fit to the observed
   CMD. In the theoretical panels, stars are color coded in azure
   (0.55${\mathcal{M}_{\odot}}$) and red
   (0.46${\mathcal{M}_{\odot}}$).
    \textit{Right panels}: one of the observed 2CDs of Fig.~\ref{f2}
    (top) and a theoretical counterpart (bottom).}}
\label{ftheory}
\end{figure}

There are a few possible theoretical explanations for the split of the
bright WDCS of \oc: (1) unresolved WD+WD binaries, (2) a combination
of DA+DB objects, or (3) bimodal WD mass distribution, presumably
related to the He-enhanced subpopulation.

Regarding the binary scenario, we note that the split WDCS at these
bright magnitudes cannot be explained by unresolved $\sim 0.54
{\mathcal{M}_{\odot}}$ WD+WD binaries.  Although the split we observe
is very nearly equal to that expected for equal-mass binaries, the
masses of the two components would have to be nearly identical at
birth, since only an extremely narrow range of initial mass occupies
the transient cooling phase at the same time.  If the mass ratio of
the progenitors is distributed approximately homogeneously between 0
and 1 (as found by Milone et al.\ 2012), then the two components would
be continuously distributed over a large magnitude range, and the net
effect would be a moderate widening of the bright WDCS , not a split
(as shown for example by the simulations for NGC~6791, Bedin et
al.~2008b).

We now explore the possibilities that the split is caused by either a
combination of DA+DB objects or a WD mass dicothomy. The top-left
panel of Fig.~\ref{ftheory} superposes on the observed WDCS four
theoretical cooling sequences: a $0.55 {\mathcal{M}_{\odot}}$ CO-core
DA track (blue solid line), the DB counterpart (blue dash-dotted
line), a $0.46 {\mathcal{M}_{\odot}}$ CO-core DA track (red dotted
line), and a He-core track (and DA atmosphere) with the same $0.46
{\mathcal{M}_{\odot}}$ mass (red dashed line).

The $0.55 {\mathcal{M}_{\odot}}$ DA model that fits the bWD sequence
for the eclipsing binary distance represents the ``canonical'' WD
population.  The DB track for the same mass is very close to the DA
track in the region of interest, so the split cannot be explained by
DA and DB populations with the same mass.  We note that a bimodal
H-layer mass distribution is also ruled out because any sequence with
lower thickness (the ``standard'' value ${\rm
  10^{-4}\mathcal{M}_{WD}}$ employed in our models is an upper
evolutionary limit) is shifted to bluer colors and cannot explain the
red sequence.

The only explanation for the rWD objects is therefore a sequence of
lower-mass WDs, which would conseguently have larger radii.  We find
that the rWD sequence can be reproduced by $\sim 0.46
\mathcal{M}_{\odot}$ models, with either CO- or He-cores.

The bottom left-hand panel of Fig.~\ref{ftheory} shows a CMD obtained
from a composite synthetic sample made of the $0.55
{\mathcal{M}_{\odot}}$ DA and $\sim 0.46 \mathcal{M}_{\odot}$ He-core
models.  The photometric error has been implemented following the
results of the data reduction.  The synthetic CMD reproduces the
bright split sequence, which in the simulation merges around $m_{\rm
  F275W} \sim 23.5$ --consistent with the observations-- due to the
increased photometric error.  The right-hand panels of
Fig.~\ref{ftheory} show one of the 2CDs of Fig.~\ref{f2}, both
observed (top) and simulated (bottom).  Notice that, like the
observations, the simulated sample also does not show the bWD+rWD
split in the color-color plane.

\section{The Origin of the rWD Sequence}
\label{sec:5}

It is tempting to map the two WD populations to the two main MS
components, and associate the rWD sequence with the helium-rich MS
population.  To test this hypothesis, we can also use the ratio of bWD
to rWD stars, and compare it with theoretical predictions.  First of
all, we need to know what kind of WDs each is made up of, since CO and
He WDs cool at different rates. An answer to these questions requires
a brief examination of the cluster HB morphology.

Cassisi et al.\ (2009, C09) showed that the progeny of the He-enhanced
MS (${\rm Y}=0.40$) must be confined to the blue tail of the observed
HB, and found that the blue clump of stars at the end of the extended
HB tail (see Fig.~2 of C09) can be reproduced by either hot flashers
or canonical He-rich HB stars, or a mixture of both types.

The minimum mass for the He-rich component along the HB is $\sim 0.46
{\mathcal{M}_{\odot}}$ (C09). These objects will form CO-core WDs with
essentially the same mass.  Even a small amount of additional mass
loss along the red giant branch (RGB) would prevent core He-ignition
and produce $\sim 0.46 \mathcal{M}_{\odot}$ He-core WDs.  Note that
these stars with high initial He abundances need to lose only $\sim$
0.10--0.15 ${\mathcal{M}_{\odot}}$ along the RGB to fall below the
limit for He-ignition, since their TO mass is about $\sim 0.2
{\mathcal{M}_{\odot}}$ less than that of the normal-He stars.

The presence of the ``mass gap'' between the two WD sequences (0.55
vs.\ 0.46 ${\mathcal{M}_\odot}$) supports the idea that the blue clump
in the HB is populated mainly by the evolved members of the He-rich
subpopulation, which eventually become CO and --possibly-- He-core WDs
with a mass of $\sim 0.46 \mathcal{M}_{\odot}$, creating the rWD
sequence.  On the other hand, the bWD sequence is instead produced by
the end-products of He-normal stars (${\rm Y}=0.25$) that populate the
rest of the observed HB. Although a fraction of the He-normal
population {\it could\,} explain the red end of the blue clump, such
stars would have $\sim 0.49 {\mathcal{M}_{\odot}}$ cores and the
resulting WDs, with essentially the same mass (because of the
vanishingly small envelopes of their HB progenitors), would occupy the
gap between the observed sequences.

With this understanding of the evolutionary origin of the rWD
sequence, we investigate whether our interpretation can also explain
the observed $N_{\rm rWD}/N_{\rm bWD}=1.8\pm0.2$ number ratio in the
magnitude range $20.4<m_{\rm F275W}<23.4$.  Our CMDs show that the
ratio between blue-clump stars and the rest of the HB population is
$\sim 0.32$, and the ratio between the He-normal and He-rich
populations along the MS is $\sim 1$ in this central field.

We consider two 13.5 Gyr isochrones from the BaSTI database
(Pietrinferni et al.~2006):\ one with [Fe/H]=$-1.6$, Y=0.246 to
represent the He-normal population, and the other with [Fe/H]=$-1.3$,
Y=0.40 to represent the He-rich component. The exact age is not
critical:\ if the age of both components is decreased by 1-2 Gyr
and/or the He-rich population is made younger by $\sim 1$~Gyr, the
quantitative result of our analysis is largely unchanged.  The
isochrones allow us to calculate the so-called {\sl evolutionary
  flux}, which tells us the number of stars leaving the MS per year:
$$
b(t)=\Phi(\mathcal{M}_{\rm TO}) \left|\frac{\rm d}{{\rm d}t} \mathcal{M}_{\rm TO}\right| = 
{\rm A}\times\mathcal{M}_{\rm TO}^{-2.3}\left|\frac{\rm d}{{\rm d}t}\mathcal{M}_{\rm TO}\right|
$$ 
(Renzini \& Buzzoni~1986). The adopted mass function ${\rm
  \Phi(\mathcal{M}_{\rm TO})}$ has the Salpeter slope.  The TO mass
${\mathcal{M}_{\rm TO}}$ and its time-derivative are determined from
the isochrones, and the constant A is fixed by constraining that the
number of MS stars (below the TO) of the two components to be the
same.  (Note that the MS is the only region that allows us to
determine the mass-function normalization and initial fraction of
He-rich stars).  The value of $b(t)$ of the He-enhanced population
turns out to be 1.1 times the value for the He-normal component. Given
that beyond the TO (apart from the faint end of the WDCS) the initial
value of the evolving mass along an isochrone is
$\sim{\mathcal{M}_{\rm TO}}$, the number of stars in a given post-MS
phase can be approximated by $N_{\rm PMS}= b(t) \times t_{\rm PMS}$
(where ${t_{\rm PMS}}$ is the lifetime of the post-MS evolutionary
stage).

The typical HB mass for stars in the He-normal component (i.e., not in
the blue clump) is $\sim$ 0.61--0.62 ${\mathcal{M}_{\odot}}$. Using
$b(t)$, the model HB timescales provide an expected number ratio of
$\sim 1.7$ between blue-clump stars (He-rich) and the rest of the HB
population.  The observed ratio is $\sim 0.32$, thus only $\sim 20$~\%
of the He-rich population must have produced blue-clump stars. This
means that $\sim 80$~\% of the He-burning stars that we expect from the
He-rich component seen along the MS, must have missed the He-ignition
and evolved onto the WDCS, as He-core WDs.  If these stars left the
RGB close to He-ignition (where mass loss is expected to be more
efficient), they now populate the rWD sequence as $\sim 0.46
{\mathcal{M}_\odot}$ He-core WDs.  Likewise, the progeny on the
(He-rich) HB blue clump produced $\sim 0.46 {\mathcal{M}_\odot}$
CO-core WDs, and the He-normal population produced ``canonical'' $0.55
\mathcal{M}_\odot$ CO-core WDs.

We can now split the {\sl evolutionary flux} for the He-rich
population into two components, according to the fraction of stars
that achieve He-ignition.  Over the magnitude range of interest, the
$0.46 {\mathcal{M}_\odot}$ CO-core model has approximately the same
cooling time of the $0.55 {\mathcal{M}_\odot}$ model (on the order of
$10^8$ yrs), while the $0.46 {\mathcal{M}_\odot}$ He-core model has
cooling timescales a factor of $\sim 2.4$ longer.  By considering
these $b(t)$ and timescale ratios, we derive an expected number ratio:

\begin{eqnarray}
\frac{N_{\rm rWD}}{N_{\rm bWD}}&=&
\frac{b(\rm prog.\ rWD)^{\rm He\ core} \times  {\it t}_{\rm rWD}^{\rm He\ core}}
{b({\rm prog.\ bWD}) \times  t_{\rm bWD}} \nonumber \\
&+&
\frac{b({\rm prog.\ rWD})^{\rm CO\ core} \times  t_{\rm rWD}^{\rm CO\ core}}
{b({\rm prog.\ bWD}) \times  t_{\rm bWD}} \nonumber 
\end{eqnarray}
and, since $t_{\rm rWD}^{\rm He\ core} / t_{\rm bWD} = 2.4$ and
$t_{\rm rWD}^{\rm CO\ core} / t_{\rm bWD} = 1.0$, we have
$$
\frac{N_{\rm rWD}}{N_{\rm bWD}}\simeq
 0.8\times1.1\times2.4 + 0.2\times 1.1 \times 1.0  \simeq 2.3, 
$$
not far from the observed value of $1.8 \pm 0.2$.  This also implies
that the number of He-core WDs along the rWD sequence is $\simeq
(0.8\times 1.1 \times 2.4) / (0.2 \times 1.1 \times 1.0) \simeq 10$
times the number of CO-core WDs.

To summarize, when taking into account the various evolutionary paths
during the post-MS phases of the He-rich and He-normal MS components
and the appropriate timescales, we can reasonably reproduce the
observed $N_{\rm rWD}/N_{\rm bWD}$ ratio, and derive a very high
fraction of He-core WDs along the red CS.

\section{Summary and Conclusions}
\label{sec:6}

Our high-precision, empirical-PSF-based photometric techniques have
revealed a never-before-seen split in the bright part of the WDCS in a
Galactic GC.  We have interpreted the observed WDCS as follows:
\begin{itemize}
\item{The blue sequence corresponds to $\sim 0.55
  {\mathcal{M}_{\odot}}$ CO-core WDs from the He-normal MS population,
  and the red sequence corresponds to $\sim 0.46
  {\mathcal{M}_{\odot}}$ both CO-core and He-core WDs from the
  He-enriched population.  The radial gradients of the various
  populations are consistent with this scenario.}
\item{The ``mass gap'' between the two WD sequences means that the
  blue clump along the cluster HB has to be populated mainly by stars
  belonging to the He-rich component.}
\item{The observed number ratio between blue-clump HB stars and the
  rest of the HB population is consistent with the number ratio of
  He-rich to He-normal MS stars, only if $\sim 80$~\% of the HB stars
  expected to be produced by the He-rich component failed to achieve
  He-ignition and evolved onto the red WDCS as $\sim 0.46
  {\mathcal{M}_{\odot}}$ He-core WDs.}
\item{It is {\it not\,} extreme mass loss that produces these He-core
  WDs, but rather the fact that the TO mass of the He-rich
  subpopulation is $\sim 0.2 {\mathcal{M}_{\odot}}$ lower than that of
  the He-normal component. A total mass loss of just $\sim 0.15
  {\mathcal{M}_{\odot}}$ along the RGB suffices to prevent central
  He-ignition for this subpopulation.}
\item{The number ratio between CO-core and He-core WDs (with longer
  cooling times) along the red WDCS is $\sim 0.1$.  This mixture of
  WDs along the red WDCS approximately explains the observed number
  ratio (rWD/bWD).  This large contribution of massive He-core WDs
  confirms the more indirect inference by Calamida et al.~(2008), that
  the bright WD sequence must be populated by a substantial fraction
  of He-core WDs.}
\end{itemize}

In conclusion, our observations and theoretical analysis have for the
first time disclosed a connection between globular clusters with
extreme blue HB stars and enhanced He, and the morphology -- hence
mass distribution -- of their WDCS. It would be extremely important to
study the bright WDCS in other clusters with extreme HBs and He-rich
subpopulations (i.e. NGC~6752, NGC~2808) to see whether a split WDCS
analogous to the one detected in \oc\ can be found.

 \acknowledgments \noindent \textbf{Acknowledgments.} AB and JA
 acknowledge support from STScI grant AR-12656.  GP acknowledges 
 partial support by the Universit\`a degli Studi di Padova CPDA101477
 grant.

\small{
}

\end{document}